\documentclass[preprint,1p]{elsarticle}
\usepackage{amsmath,amssymb,amsfonts}
\usepackage{bbm}
\usepackage{mathrsfs}
\usepackage{algorithm}
\usepackage{algorithmic}
\usepackage{color}
\usepackage{graphicx}
\usepackage{natbib}
\usepackage[pdftex,colorlinks=true]{hyperref}

\providecommand{\ud}{\,\mathrm{d}}
\providecommand{\abs}[1]{\left\lvert#1\right\rvert}
\providecommand{\bra}[1]{\left\langle#1\right\vert}
\providecommand{\ket}[1]{\left\vert#1\right\rangle}
\providecommand{\hprod}[2]{\left\langle#1\,\vert\,#2\right\rangle}
\providecommand{\bprod}[2]{\left(#1\,\vert\,#2\right)}
\providecommand{\vect}[1]{{\boldsymbol{#1}}}

\providecommand{\suchthat}{\,:\,}

\providecommand{\conj}[1]{\overline{#1}}

\providecommand{\acom}[2]{\left\{#1,#2\right\}}
\providecommand{\trace}{\mathrm{tr}}

\providecommand{\R}{{\mathbb R}}
\providecommand{\C}{{\mathbb C}}

\providecommand{\N}{{\mathbb N}}

\providecommand{\bitand}{\land}
\providecommand{\bitor}{\lor}
\providecommand{\bitxor}{\oplus}
\providecommand{\bitshiftleft}{\ll}
\providecommand{\bitshiftright}{\gg}
\providecommand{\lastbit}[1]{\mathrm{LastBit}\!\left(#1\right)}
\providecommand{\bitcount}[1]{{\#\!\left(#1\right)}}
\providecommand{\bitlabel}[2]{\stackrel{{\color{blue}#1}}{{#2}}}
\providecommand{\revp}[1]{{\sigma_\mathrm{revperm}\left(#1\right)}}

\definecolor{med-gray}{gray}{0.5}

\algsetup{linenosize=\tiny}

\journal{Computer Physics Communications}

\begin{document}

\begin{frontmatter}

\title{The FermiFab Toolbox for Fermionic\\Many-Particle Quantum Systems}

\author{Christian B. Mendl}
\ead{christian\_mendl@hotmail.com}
\address{Center for Mathematics M7, TU M\"unchen, Germany}

\begin{abstract}
This paper introduces the \texttt{FermiFab} toolbox for many-particle quantum systems. It is mainly concerned with the representation of (symbolic) fermionic wavefunctions and the calculation of corresponding reduced density matrices (RDMs). The toolbox transparently handles the inherent antisymmetrization of wavefunctions and incorporates the creation/annihilation formalism. Thus, it aims at providing a solid base for a broad audience to use fermionic wavefunctions with the same ease as matrices in Matlab, say. Leveraging symbolic computation, the toolbox can greatly simply tedious pen-and-paper calculations for concrete quantum mechanical systems, and serves as ``sandbox'' for theoretical hypothesis testing. \texttt{FermiFab} (including full source code) is freely available as a plugin for both Matlab and Mathematica.
\end{abstract}


\end{frontmatter}

\section*{Program Summary}

\noindent\emph{Program title:} FermiFab\\
\emph{Program author:} Christian B. Mendl\\
\emph{Distribution format:} tar.gz, zip\\
\emph{No. of bytes in distributed program, including test data, etc.:} $16.3 \cdot 10^6$\\
\emph{No. of lines in distributed program, including test data, etc.:} $\sim 8000$\\
\emph{Programming language:} MATLAB, Mathematica, C\\
\emph{Computer:} PCs, Sun Solaris workstation\\
\emph{Operating system:} any platform supporting MATLAB or Mathematica; tested with Windows (32 and 64 bit) and Sun Solaris\\
\emph{RAM:} case dependent\\

\section{Introduction}
\label{sec:Introduction}

The ground state energy of fermionic many-particle quantum systems can be re-expressed as a linear functional of (one- or two-body) reduced density matrices (RDMs). This notion traces back to the origins of quantum mechanics~\cite{Neumann1932,Dirac1930} around 1930. Since 1964, the one-body RDM has been greatly popularized by density functional theory~\cite{HohenbergKohn1964,KohnSham1965}, which is typically the most viable approximation for handling large particle numbers. The tantalizing possibility of employing RDMs (instead of many-particle wavefunctions) for \emph{exact} groundstate energy computations is counterbalanced by the $N$-representability problem, i.e., the search for necessary and sufficient conditions a two-body density must obey to represent an $N$-electron wavefunction~\cite{Loewdin1955,Coleman1963,Ando1963}. Modern applications use variational principles and semidefinite programming to impose positivity constraints on the two-body RDM~\cite{Mazziotti2007}. In any case, it is desirable to render the powerful RDM framework accessible to a broader audience, integrating it into the symbolic language of modern computer algebra systems like Mathematica, or numeric software like Matlab.

The \texttt{FermiFab} toolbox (available for download at~\cite{FermiFabSourceforge}) is precisely designed for that purpose. A short ``usage manual'' and a brief tour of the essential features is provided in the following subsections. Note that the underlying one-particle orbitals (see below) are always assumed to be orthonormalized. In addition, the toolbox adheres to the trace-normalization convention $\trace_{\wedge^p \mathcal{H}} \gamma_{\ket{\psi}\bra{\psi}} = \binom{N}{p}$ for the $p$-body RDM $\gamma_{\ket{\psi}\bra{\psi}}$ of a normalized $N$-body wavefunction $\psi$. Here, $\wedge^p \mathcal{H}$ denotes the $p$-particle Fock-space (see following subsection).

\subsection{Fermi states} 
\begin{figure}[ht!]
\begin{center}
\includegraphics[width=0.6\textwidth]{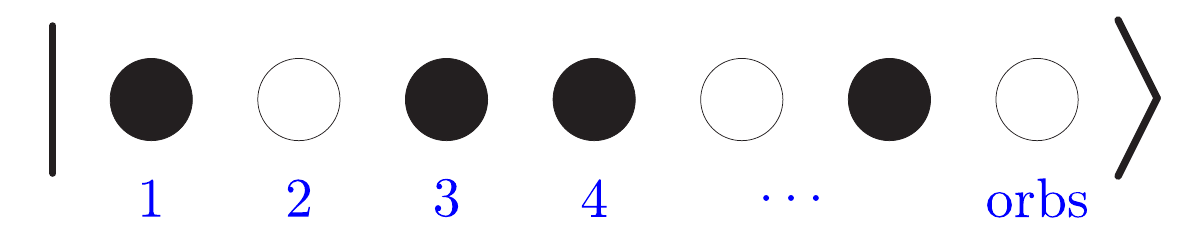}
\end{center}
\caption{Schematic illustration of a Slater determinant: (filled) circles correspond to (occupied) orbitals.}
\label{fig:Slater}
\end{figure}

Fundamental building blocks of multi-fermion quantum systems are Slater determinants (figure~\ref{fig:Slater}). These can be thought of as a collection of ``orbitals'' (or slots), some of which are occupied by a fermionic particle (e.g., an electron).
In mathematical terms, the available number of orbitals '$\mathrm{orbs}$' is the dimension of the underlying one-particle Hilbert space $\mathcal{H}$ and the number of occupied orbitals the particle number $N$. Thus there are altogether $\binom{\mathrm{orbs}}{N}$ Slater determinants. Their complex span defines the $N$-particle \emph{Fock-space} $\wedge^N \mathcal{H}$. The $N$-particle Fermi states are precisely the elements of $\wedge^N \mathcal{H}$.

\subsection{Getting started with \texttt{FermiFab}}
For concreteness, the following examples are issued in the Matlab programming language. (The Mathematica version of \texttt{FermiFab} provides the same features; section~\ref{sec:Application} contains a demonstration.) Commands typed by the user are preceded by \texttt{>>}, and the subsequent lines show the corresponding output. In standard Matlab syntax, \texttt{zeros(n,1)} below constructs a column vector of length $n$, and \texttt{nchoosek} computes binomial coefficients. We first show how to represent an $N = 4$ particle state $\psi$ with, e.g., $6$ available orbitals in total:
{\small\begin{verbatim}
>> orbs = 6; N = 4;
>> x = zeros(nchoosek(orbs,N),1); x(1)=1/sqrt(2); x(2)=1i/sqrt(2);
>> psi = fermistate(orbs,N,x)

psi = 
	Fermi State (orbs == 6, N == 4)
	(0.70711)|1234> + (0+0.70711i)|1235>
\end{verbatim}}
\noindent Needless to say, the \texttt{fermistate} command is specific to the \texttt{FermiFab} toolbox. The vector $x$ contains the Slater determinant coefficients of $\psi$ in lexicographical order. Let's assign more meaningful names to the orbitals of $\psi$:
{\small\begin{verbatim}
>> psi = set(psi,'orbnames',{'1s' '1s~' '2s' '2s~' '2p' '2p~'})

psi = 
	Fermi State (orbs == 6, N == 4)
	(0.70711)|1s 1s~ 2s 2s~> + (0+0.70711i)|1s 1s~ 2s 2p>
\end{verbatim}}
\noindent From a physics viewpoint, these orbitals could form electronic subshells in atoms. The rank-one projector $\ket{\psi}\bra{\psi}$ or ``density matrix'' of $\psi$ can be calculated intuitively by
{\small\begin{verbatim}
>> psi*psi'

ans = 
	Fermi Operator wedge^4 H -> wedge^4 H (orbs == 6)

Matrix representation w.r.t. ordered Slater basis
(|1s 1s~ 2s 2s~>, |1s 1s~ 2s 2p>, ... |2s 2s~ 2p 2p~>) ->
    (|1s 1s~ 2s 2s~>, |1s 1s~ 2s 2p>, ... |2s 2s~ 2p 2p~>):

  Columns 1 through 4

   0.5000                  0 - 0.5000i        0                  0
        0 + 0.5000i   0.5000                  0                  0
   ...
\end{verbatim}}
\noindent Note that the result is now a \texttt{fermiop} operator acting on $\wedge^4 \mathcal{H}$.

\subsection{Reduced density matrices} The core feature of the toolbox is the efficient calculation of RDMs. For example, the 2-body RDM
\begin{equation*}
\hprod{ij}{\gamma_{\ket{\psi}\bra{\psi}}\,kl} := \hprod{\psi}{a^\dagger_k a^\dagger_l a_j a_i\,\vert\,\psi}
\end{equation*}
can be obtained by
{\small\begin{verbatim}
>> rdm(psi,2)

ans = 
	Fermi Operator wedge^2 H -> wedge^2 H (orbs == 6)

Matrix representation w.r.t. ordered Slater basis
(|1s 1s~>, |1s 2s>, ... |2p 2p~>) -> (|1s 1s~>, |1s 2s>, ... |2p 2p~>):

  Columns 1 through 4

   1.0000                  0                  0                  0
        0             1.0000                  0                  0
        0                  0             0.5000                  0 - 0.5000i
        0                  0                  0 + 0.5000i   0.5000
   ...
\end{verbatim}}
RDMs are reviewed in more detail in section~\ref{sec:ImplementationRDMs}.


\subsection{Tensor products of operators}

Given a linear operator $A: \mathcal{H} \to \mathcal{H}$, a straightforward derivation based on the antisymmetrized structure of $\wedge^N \mathcal{H}$ shows that
\begin{equation*}
\hprod{j_1,\dots,j_N}{(A \otimes\cdots\otimes A) \,\vert\, i_1,\dots,i_N} = \det \hprod{j_k}{A\,\vert\,i_\ell}_{k,\ell}
\end{equation*}
for all $1\le i_1 < \dots < i_N \le \dim\mathcal{H}$ and $1\le j_1 < \dots < j_N \le \dim\mathcal{H}$. That is, we obtain a matrix representation of $A \otimes\cdots\otimes A$ acting on $\wedge^N \mathcal{H}$. The \texttt{tensor\_op} command implements precisely this operation. The following code lines are taken from the ``natural orbitals'' example in \texttt{test/norbs.m}:
{\small\begin{verbatim}
>> orbs = 6; N = 4;
>> psi = fermistate(orbs,N,crand(nchoosek(orbs,N),1));
>> [U,D] = eig(rdm(psi,1));
\end{verbatim}}
\noindent \texttt{crand} generates pseudorandom complex numbers (similar to \texttt{rand}), and \texttt{eig} computes eigenvalues and -vectors. Thus, the eigenvectors of the 1-body RDM of $\psi$ are stored in $U$. Performing a corresponding base change on $\wedge^N \mathcal{H}$ using these eigenvectors should result in a diagonal 1-body RDM~\cite{Loewdin1955}:
{\small\begin{verbatim}
>> psi = tensor_op(U,N)'*psi;
>> G = get(rdm(psi,1),'data');
>> err = norm(G-diag(diag(G)))

err =

  1.6512e-015
\end{verbatim}}
\noindent In many physical applications, one can take advantage of unitary base changes on $\mathcal{H}$ such that subsequent computations are simplified, e.g., by choosing single-particle eigenstates of the $L_z$ angular momentum operator. The above code shows how to implement the according base change on $\wedge^N \mathcal{H}$.

\subsection{State configurations} For performance and memory efficiency reasons, \texttt{FermiFab} has built-in ``configurations'', i.e., we can subdivide the available orbitals into several groups, each of which contains a fixed number of particles. (Physically speaking, the groups could be interpreted as atomic subshells $1s,2s,2p,3s$, for example.) Let's say our system involves a total of 3 particles in 9 orbitals, with exactly 2 particles in the first 5 orbitals and 1 particle in the remaining 4 orbitals. Then a \texttt{fermistate} reflecting this configuration is specified by
{\small\begin{verbatim}
>> orbs = [5,4]; N = [2,1];
>> psi = fermistate(orbs,N)

psi = 
	Fermi State (orbs == 9, N == 3)
	|126>
\end{verbatim}}
\noindent Note that $\ket{126}$ is the lexicographically first base vector respecting the configuration constraints, and that $\psi$ requires only $\binom{5}{2} \cdot \binom{4}{1} = 40$ rather than $\binom{9}{3} = 84$ complex entries:
{\small\begin{verbatim}
>> length(psi)

ans =

    40
\end{verbatim}}
\noindent The \texttt{rdm} command works transparently for any configuration, so $\psi$ behaves like a standard 9-orbital, 3-particle state.

What happens if we add two \texttt{fermistate}s with different but compatible configurations (i.e., the total number of orbitals and particles is the same)?
{\small\begin{verbatim}
>> orbs = [2,7]; N = [1,2];
>> phi = fermistate(orbs,N)

phi = 
	Fermi State (orbs == 9, N == 3)
	|134>

>> length(phi)

ans =

    42

>> chi = psi+phi

chi = 
	Fermi State (orbs == 9, N == 3)
	|126> + |134>
\end{verbatim}}
\noindent as expected -- so how is this accomplished? \texttt{FermiFab} has detected that it needs to combine the two configurations, resulting in the full-fledged 9-orbital, 3-particle state. This fact can be checked by
{\small\begin{verbatim}
>> length(chi)

ans =

    84
\end{verbatim}}

\subsection{Symbolic computations} The Mathematica version of \texttt{FermiFab} is -- quite naturally -- inherently based on symbolic language. Considering Matlab, the (optionally available) Symbolic Math Toolbox integrates seamlessly into \texttt{FermiFab}, too. Taking advantage of symbolic computations is thus easily accomplished. That is, in the above examples, we may as well insert symbolic variables:
{\small\begin{verbatim}
>> syms a b c
>> y = sym(zeros(1,nchoosek(orbs,N)));
>> y(1) = a; y(3) = 1i*b^2; y(4) = 1/c;
>> psi = set(psi,'data',y)

psi = 
	Fermi State (orbs == 6, N == 4)
	(a)|1s 1s~ 2s 2s~> + (b^2*i)|1s 1s~ 2s 2p~> + (1/c)|1s 1s~ 2s~ 2p>

>> rdm(psi,2)

ans = 
	Fermi Operator wedge^2 H -> wedge^2 H (orbs == 6)

Matrix representation w.r.t. ordered Slater basis
(|12>, |13>, ... |56>) -> (|12>, |13>, ... |56>):

[ (c*b^2*conj(b)^2 + a*c*conj(a))/c + 1/(c*conj(c)), 
...
\end{verbatim}}

\section{Implementation Details}
\label{sec:Implementation}


The algorithmic implementation is based on the canonical mapping from Slater determinants to \emph{bitfields}. That is, each Slater determinant corresponds to an unsigned integer $s$, where the $i$th bit is set to 1 precisely when the $i$th orbital is occupied. To remain unambiguous in terms of bitlength, the first orbital is stored in the LSB (least significant bit). Now, our task consists of re-expressing the creation/annihilation and RDM formalism in terms of bit operations. Note that, for example, testing whether all occupied orbitals in $s_1$ are also occupied in $s_2$ amounts to the pretty simple line of code $s_1 \bitand s_2 = s_1$, where we have used the bitwise AND operator $\bitand$. The following table summarizes all required bit operations:\\
\begin{center}
\begin{tabular}{ll}
bitwise AND:&$x \bitand y$\\
bitwise OR:&$x \bitor y$\\
bitwise XOR:&$x \bitxor y$\\
bit shift left:&$x \bitshiftleft n$\\
bit shift right:&$x \bitshiftright n$\\
bit count:&$\bitcount{x}$
\end{tabular}
\end{center}
\medskip
\noindent For example, $1001101_2 \bitshiftright 3 = 1001_2$ and $\bitcount{18} = \bitcount{10010_2} = 2$. Note that bit operations are typically very ``cheap'' on CPUs. (In particular, refer to the SSE4~\cite{SSE4} \texttt{POPCNT} ``population count'' instruction for bit counting.) Diving a little bit further down into CPU intrinsics, we will make use of two's-complement arithmetic for negating numbers~\cite{Intel2010Architecture}, e.g.,
\begin{equation}
\label{eq:TwosComplement}
\begin{split}
 x &= \dots 001011{\color{red}1}00_2 \quad\leadsto\\
-x &= \dots 110100{\color{red}1}00_2.
\end{split}
\end{equation}
Interestingly, precisely all bits flip which are more significant than the least significant 1-bit (marked red). Thus, we can use this property to extract the last 1-bit from a bitfield $x \neq 0$ simply by
\begin{equation*}
\lastbit{x} := x \bitand (-x).
\end{equation*}
(An less universal alternative is the \texttt{BSF} ``bit scan forward'' instruction~\cite{Intel2010InstructionSet}, which returns the index of the least significant 1-bit.)

\subsection{Enumerating Slater determinants}

The basic task we set out to accomplish in this subsection is lexicographically enumerating all Slater determinants of a fixed particle number $N$ and number of orbitals '$\mathrm{orbs}$'. This amounts to computing the lexicographically next bit permutation (denoted by '$\mathrm{NextFermi}$'). For example,
\begin{equation*}
\begin{split}
s &= 0{\color{red}1}111000_2 \quad\leadsto\\
\mathrm{NextFermi}(s) &= {\color{red}1}0000111_2.
\end{split}
\end{equation*}
Closer inspection reveals the general rule that the leading 1-bit (marked red) in the least significant block of 1s gets shifted to the left by one position, and the remaining 1-bits are shifted to the end. Algorithm~\ref{alg:NextFermi} is adopted from~\cite{BitTwiddling} and performs exactly this computation. In line~\ref{algline:trailingzeros}, $s \bitor (s - 1)$ sets the trailing zeros in $s$ to 1, so for example, $s = 0{\color{red}1}111000_2$ $\leadsto$ $s \bitor (s - 1) = 0{\color{red}1}111111_2$ and $t = {\color{red}1}0000000_2$. The second term in line~\ref{algline:shiftblock} performs the shifting of the remaining 1-bits to the end.
\begin{algorithm}
\caption{NextFermi}
\label{alg:NextFermi}
\begin{algorithmic}[1]
\REQUIRE $s$: \textbf{bitfield}
\STATE $t \Leftarrow (s \bitor (s - 1)) + 1$ \label{algline:trailingzeros}
\RETURN $t \bitor (((\lastbit{t} - 1) / \lastbit{s}) \bitshiftright 1)$ \label{algline:shiftblock}
\end{algorithmic}
\end{algorithm}

As an extension of Algorithm~\ref{alg:NextFermi}, we want to take into account ``configurations'', i.e., a subdivision of the available orbitals into
several groups, each of which contains a fixed number of particles. For example, we compartmentalize a total of $11$ orbitals such that exactly $4$ particles are in the first $6$ orbitals and $2$ in the remaining $5$ orbitals, written as $(\mathrm{orbs}_1,\mathrm{orbs}_2) = (6,5)$ and $(N_1,N_2) = (4,2)$. Then a sequence of patterns -- respecting the configuration restrictions -- would be
\begin{equation}
\label{eq:ConfigEnum}
\begin{split}
&0|01010|110110_2,\\
&0|01010|11{\color{red}1}00{\color{red}1}_2,\\
&0|01010|1110{\color{red}1}0_2,\\
&0|01010|111{\color{red}1}00_2,\\
&0|01{\color{red}1}00|00{\color{red}1111}_2,
\end{split}
\end{equation}
where we have highlighted the currently changing 1-bits by red colors.

More formally, given $(\mathrm{orbs}_1,\dots,\mathrm{orbs}_k)$, the compartmentalization may be written as $V_j := \mathrm{span}\left\{\ket{i} \suchthat b_{j-1} < i \le b_j\right\} \subset \mathcal{H}$ with $b_j := \sum_{\ell=1}^j\mathrm{orbs}_\ell$. In other words, $\mathcal{H} = \bigoplus_j V_j$. In the example above, $V_1 = \mathrm{span}\{\ket{1},\dots,\ket{6}\}$ and $V_2 = \mathrm{span}\{\ket{7},\dots,\ket{11}\}$. Now, mathematically speaking, a \emph{configuration} of an $N$-particle state is a subspace of $\wedge^N\mathcal{H}$ of the following form:
\begin{equation}
\label{eq:ConfigDef}
\mathcal{C}^{N_1,\dots,N_k} := \mathrm{span}\left\{\ket{i_1,\dots,i_N} \suchthat \sharp\left\{\ell \suchthat \ket{i_\ell} \in V_j\right\} = N_j\right\}
\end{equation}
where $(N_1,\dots,N_k)$ is a partition of $N$ (i.e.\ $0\le N_j \le \mathrm{orbs}_j,\, \sum_j N_j = N$). A quantum chemist could interpret the $V_j$ as atomic subshells $1s,2s,2p,3s,\dots$ and the $N_j$ as occupation numbers. An interesting consequence of definition~\eqref{eq:ConfigDef} is the recovery of a tensor product structure, namely
\begin{equation}
\label{eq:ConfigTensor}
\mathcal{C}^{N_1,\dots,N_k} \cong \bigotimes_{j=1}^k \wedge^{N_j} V_j.
\end{equation}
This follows from the observation that a configuration is constructed by the lexicographical enumeration of Slater determinants within orbital groups, as illustrated in~\eqref{eq:ConfigEnum}.

Algorithm~\ref{alg:NextFermiConfig} implements precisely this enumeration. In accordance with the lexicographical scheme, it first iterates through all Slater determinants within the least significant orbital group (line~\ref{algline:nextlocalfermi}), then resets this group (first term in line~\ref{algline:reset}) and recursively computes the next bit pattern for the remaining groups (line~\ref{algline:recursivenext}). The $\mathrm{mask}$ in line~\ref{algline:groupmask} is required for testing whether the last bit permutation within the least significant group has been reached (line~\ref{algline:groupdone}). In the example above, we would have $\mathrm{mask} = 0|111111_2$.
\begin{algorithm}
\caption{NextFermiConfig}
\label{alg:NextFermiConfig}
\begin{algorithmic}[1]
\REQUIRE $s$: \textbf{bitfield}, orbs: \textbf{int array}
\STATE $\mathrm{mask} \Leftarrow (1 \bitshiftleft \mathrm{orbs}[0]) - 1$ \label{algline:groupmask}
\IF{$(((s \bitor (s - 1)) \bitand \mathrm{mask}) \neq \mathrm{mask})$} \label{algline:groupdone}
	\RETURN $\mathrm{NextFermi}(s)$ \label{algline:nextlocalfermi}
\ELSE
	\IF{orbs.length = 1}
		\RETURN -1
	\ENDIF
	\STATE $t \Leftarrow \mathrm{NextFermiConfig}(s \bitshiftright \mathrm{orbs}[0], \mathrm{orbs}[1,\dots,\mathrm{end}])$ \label{algline:recursivenext}
	\IF{t = -1}
		\RETURN -1
	\ENDIF
	\RETURN $(\mathrm{mask}/\lastbit{s}) \bitor (t \bitshiftleft \mathrm{orbs}[0])$ \label{algline:reset}
\ENDIF
\end{algorithmic}
\end{algorithm}

\subsection{Creation/annihilation operators}

\begin{figure}[ht!]
\begin{center}
\includegraphics[width=0.75\textwidth]{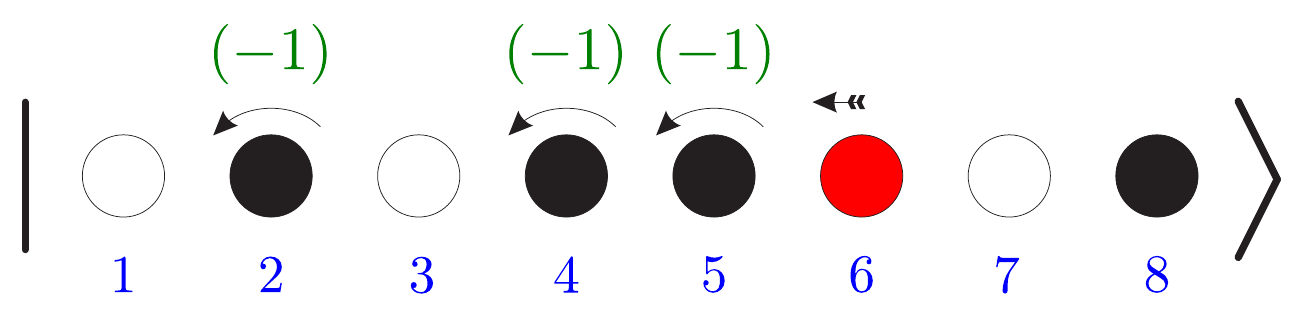}
\end{center}
\caption{Annihilation of a single orbital (red). Figuratively, the red orbital moves to the front before being removed, such that each flip (curved arrows) with an occupied preceding orbital contributes a sign factor of $(-1)$. In terms of quantum mechanics, $a_{\ket{6}} \ket{24568} = -\ket{2458}$.}
\label{fig:AnnihilSingle}
\end{figure}

The creation/annihilation operator formalism is an essential ingredient of many-particle quantum mechanics and quantum field theory~\cite{PeskinSchroeder}. For a very brief sketch, let $\varphi \in \wedge^p \mathcal{H}$ be a $p$-particle wavefunction with $1 \le p \le N$. Then, the linear \emph{annihilation operator} $a_\varphi$ acting on $\wedge^N \mathcal{H}$ removes or ``annihilates'' the state $\varphi$ from $\wedge^N \mathcal{H}$. More precisely, $a_\varphi$ is uniquely determined by its antilinearity in $\varphi$,
\begin{equation*}
a_{c\,\varphi_1 + \varphi_2} = \conj{c}\,a_{\varphi_1} + a_{\varphi_2} \quad \forall\, c\in\C, \varphi_1, \varphi_2 \in \wedge^p\mathcal{H}
\end{equation*}
together with the decomposition for Slater determinants,
\begin{equation*}
a_{\ket{i_1,i_2,\dots i_p}} := a_{\ket{i_p}} \cdots a_{\ket{i_2}} a_{\ket{i_1}} \quad \forall\, 1 \le i_1 < \dots < i_p \le \dim\mathcal{H},
\end{equation*}
as well as the definition
\begin{equation*}
a_{\ket{i}} \ket{j_1,\dots,j_N} := \left\{
\begin{array}{cl}
(-1)^{k-1} \ket{j_1,\dots,j_{k-1},j_{k+1},\dots,j_N}& i = j_k\\
0&i \notin \{j_1,\dots,j_N\}\end{array}\right.
\end{equation*}
for all $1 \le j_1 < \dots < j_N \le \dim\mathcal{H}$. The sign factor can be interpreted as number of orbital ``flips'' illustrated in figure~\ref{fig:AnnihilSingle}.


So far we have considered annihilation operators only. The \emph{creation operator} $a^\dagger_\varphi$ is by definition the adjoint (conjugate transpose) of $a_\varphi$, as the notation already suggests. It can be shown that the following relations hold, where the anticommutator bracket is defined by $\acom{A}{B} := A\,B - B\,A$ and $\varphi, \chi \in \wedge^p \mathcal{H}$ are arbitrary wavefunctions:
\begin{equation*}
\acom{a_\varphi}{a_\chi} = 0, \quad \acom{a^\dagger_\varphi}{a^\dagger_\chi} = 0, \quad \acom{a_\varphi}{a^\dagger_\chi} = \hprod{\varphi}{\chi}.
\end{equation*}

In the remainder of this subsection, we want to detail an efficient algorithmic implementation of the annihilation operation, w.l.o.g.\ for Slater determinants only. More precisely, let $\ket{s} \in \wedge^N \mathcal{H}$ be a fixed Slater determinant, then our task is the calculation of $a_{\ket{t}}\ket{s}$ for arbitrary Slater determinants $\ket{t} \in \wedge^p \mathcal{H}$ and $1 \le p \le N$. The result will be nonzero only if all occupied orbitals in $t$ are also occupied in $s$, which can be tested by $t \bitand s = t$ as already mentioned in the beginning. Given this holds true, the bit pattern describing the Slater determinant $a_{\ket{t}}\ket{s}$ is simply $s - t$, so what essentially remains is the calculation of the sign factor.


For that purpose, we define the \emph{annihilation sign mask} of $s$ such that each bit stores the integer parity of the number of less or equal significant 1-bits in $s$. That is, if $s$ has binary representation
\begin{align*}
s =& \dots a_2 a_1 a_0, \quad a_i \in \{0,1\}, \quad \text{then}\\
\mathrm{AnnihilSignMask}(s) :=& \dots b_2 b_1 b_0 \quad\text{with}\quad b_i \equiv \sum_{j=0}^i a_j \!\mod 2.
\end{align*}
For example, $s = \bitlabel{7}{0}\bitlabel{6}{0}\bitlabel{5}{1}\bitlabel{4}{0}\bitlabel{3}{1}\bitlabel{2}{1}\bitlabel{1}{0}\bitlabel{0}{0}_2$ results in $\mathrm{AnnihilSignMask}(s) = \dots\bitlabel{7}{1}\bitlabel{6}{1}\bitlabel{5}{1}\bitlabel{4}{0}\bitlabel{3}{0}\bitlabel{2}{1}\bitlabel{1}{0}\bitlabel{0}{0}_2$ where blue overhead numbers label bit positions. Algorithm~\ref{alg:AnnihilSignMask} implements this calculation. It has a running time of $\mathcal{O}(\bitcount{s})$ since the last statement (line~\ref{algline:removelastbit}) in the while loop removes the least significant 1-bit from $s$. In line~\ref{algline:addparities}, the $\bitxor (-t)$ operation flips all bits which are less or equal significant than the current least significant 1-bit.
\begin{algorithm}
\caption{AnnihilSignMask}
\label{alg:AnnihilSignMask}
\begin{algorithmic}[1]
\REQUIRE $s$: \textbf{bitfield}
\STATE \textbf{bitfield} $m \Leftarrow 0$
\WHILE{$s \neq 0$}
	\STATE $t \Leftarrow \lastbit{s}$
	\STATE $m \Leftarrow m \bitxor (-t)$ \label{algline:addparities}
	\STATE $s \Leftarrow s - t$ \label{algline:removelastbit}
\ENDWHILE
\RETURN $m$
\end{algorithmic}
\end{algorithm}

Finally, we define the \emph{reverse permutation sign} $\revp{n}$ for all $n \in \N_{\ge 1}$ by the sign of the permutation $i \mapsto n-i+1$ ($i = 1,\dots,n$). A moment's thought reveals that
\begin{equation*}
\revp{n} = (-1)^{\frac{1}{2}(n-1)n}.
\end{equation*}

\begin{figure}[ht!]
\begin{center}
\includegraphics[width=0.75\textwidth]{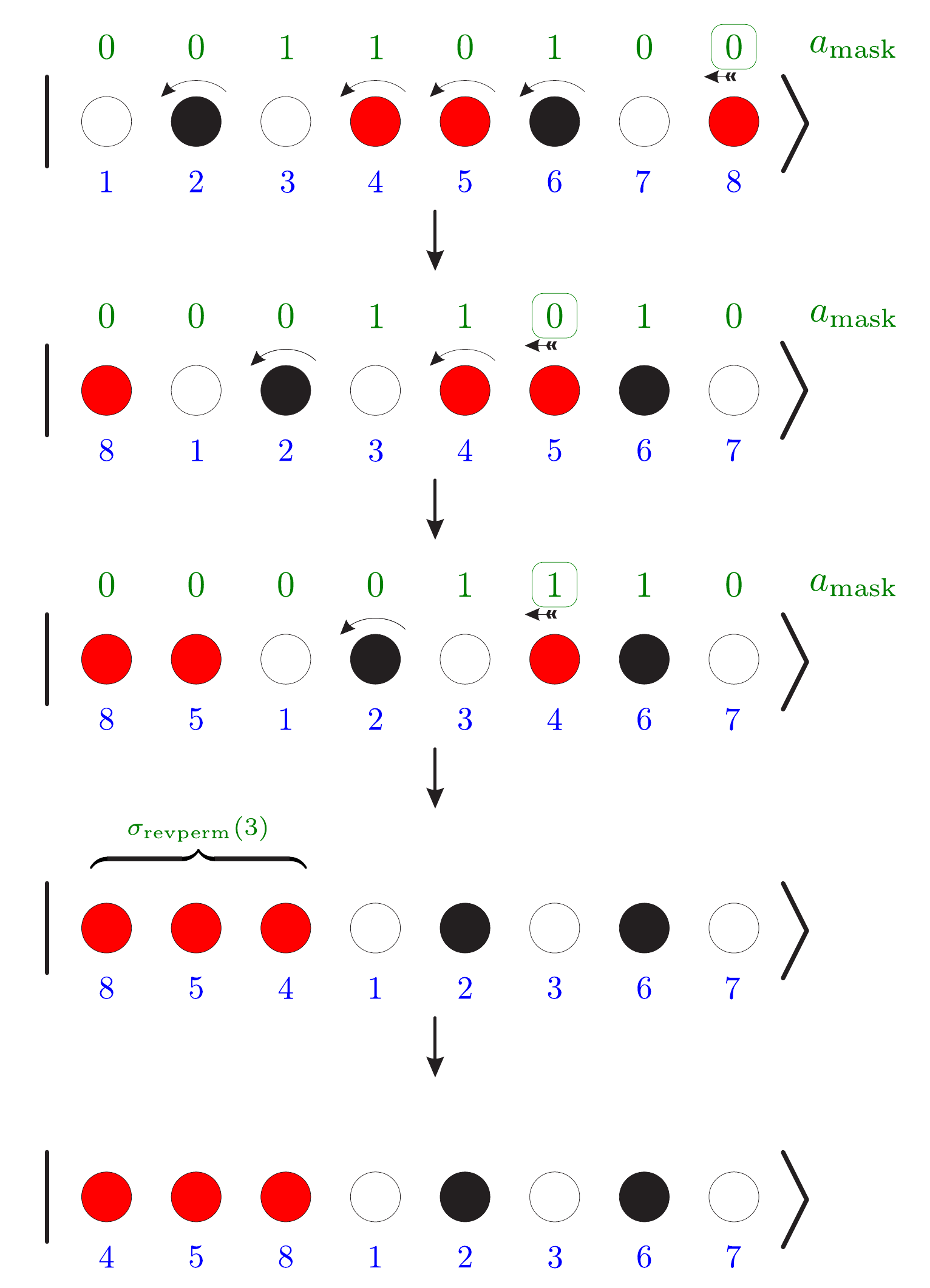}
\end{center}
\caption{The cumulative sign factors incurred during rearrangement of the to-be annihilated (red) orbitals to the front. In terms of quantum mechanics, the corresponding  operation reads $a_{\ket{458}} \ket{24568} = \ket{26}$. The contribution from all flips (curved arrows) during each step can be obtained from the marked bit in $a_\mathrm{mask}$. Note that this mask needs to be calculated only once. The permutation sign for sorting the three red orbitals in the last step equals $\revp{3} = -1$, so the overall sign factor is $1$. Algorithmically, the whole schematic is implemented by equation~\eqref{eq:AnnihilSignFactor}.}
\label{fig:AnnihilSignfactor}
\end{figure}

Altogether, our devised algorithm is illustrated in figure~\ref{fig:AnnihilSignfactor}. More formally, we obtain $a_{\ket{t}}\ket{s} = \zeta\cdot\ket{s-t}$ with the sign factor $\zeta$ equal to
\begin{equation}
\label{eq:AnnihilSignFactor}
\zeta = \revp{\bitcount{t}} \cdot (-1)^\bitcount{a_\mathrm{mask} \bitand t},
\end{equation}
where we have set $a_\mathrm{mask} := \mathrm{AnnihilSignMask}(s) \bitshiftleft 1$. Equation~\eqref{eq:AnnihilSignFactor} will be the basic building block for calculating reduced density matrices in Algorithm~\ref{alg:SlaterRDM} below, as described in the next subsection.

\subsection{Reduced density matrices}
\label{sec:ImplementationRDMs}

In this subsection we briefly recall the relevant abstract formalism, and then describe the algorithmic implementation in the \texttt{FermiFab} toolbox. Let $1 \le p_k \le N_k$ ($k = 1,2$) and denote orthonormal basis sets of $\wedge^{p_k}\mathcal{H}$ by $\left(\varphi_{ki}\right)_i$. For wavefunctions $\psi_k \in \wedge^{N_k}\mathcal{H}$ ($k = 1,2$), define the \emph{reduced density matrix} $\gamma_{\ket{\psi_1}\bra{\psi_2}}: \wedge^{p_2}\mathcal{H} \to \wedge^{p_1}\mathcal{H}$ by
\begin{equation}
\label{eq:DefRdm}
\hprod{\varphi_{1j}}{\gamma_{\ket{\psi_1}\bra{\psi_2}}\,\vert\,\varphi_{2i}} := \hprod{\psi_2}{a^\dagger_{\varphi_{2i}} a_{\varphi_{1j}} \psi_1} = \hprod{a_{\varphi_{2i}} \psi_2}{a_{\varphi_{1j}} \psi_1} \quad \forall\, i,j,
\end{equation}
where we have employed the creation/annihilation operators defined in the last subsection. The significance of this definition can be seen as follows. Any linear map $b: \wedge^{p_1}\mathcal{H} \to \wedge^{p_2}\mathcal{H}$ with matrix representation $\left(b_{ij}\right)$ may be ``lifted'' to an operator $B: \wedge^{N_1}\mathcal{H} \to \wedge^{N_2}\mathcal{H}$ by
\begin{equation}
\label{eq:p2N}
B := \sum_{i,j} b_{ij}\,a^\dagger_{\varphi_{2i}} a_{\varphi_{1j}}.
\end{equation}
(A prominent example is the Coulomb operator ($p_1 = p_2 = 2$), which describes the \emph{pairwise} interaction between charged particles.) Now, the $B$ expectation value with respect to $\ket{\psi_1}\bra{\psi_2}$ equals
\begin{equation}
\label{eq:AvrRDM}
\hprod{\psi_2}{B\,\psi_1} \stackrel{\text{def}}{=} \sum_{i,j} b_{ij} \hprod{\psi_2}{a^\dagger_{\varphi_{2i}} a_{\varphi_{1j}}\,\psi_1} = \trace_{\wedge^{p_2}
\mathcal{H}}\left(b\,\gamma_{\ket{\psi_1}\bra{\psi_2}}\right).
\end{equation}
In other words, this equation switches from $\wedge^{N_k}\mathcal{H}$ to $\wedge^{p_k}\mathcal{H}$ ($k = 1,2$). For many applications, this is the only possibility to avoid the ``curse of dimensionality'' induced by the $N_1$, $N_2$-particle systems. In terms of \texttt{FermiFab},~\eqref{eq:p2N} is implemented by the \texttt{p2N} command.

In the rest of this subsection, we focus on the calculation of $\gamma_{\ket{\psi_1}\bra{\psi_2}}$ in Algorithm~\ref{alg:SlaterRDM}. Due to linearity, it suffices to restrict ourselves to Slater determinants. That is, $\psi_1$ and $\psi_2$ are (w.l.o.g.) replaced by Slater determinants $s_1$ and $s_2$, respectively, and it is assumed that the $\left(\varphi_{ki}\right)$ are Slater determinants, too. So the last term in~\eqref{eq:DefRdm} can be concisely written as $\hprod{a_{t_2} s_2}{a_{t_1} s_1}$ with Slater determinants $t_k \in \wedge^{p_k}\mathcal{H}$ ($k = 1,2$). Note that the particle number conservation law imposes $N_1 - p_1 = N_2 - p_2$, otherwise all terms will be zero; so we calculate $p_2$ from given $N_1$, $N_2$ and $p_1$.

\begin{figure}[ht!]
\begin{center}
\includegraphics[width=0.9\textwidth]{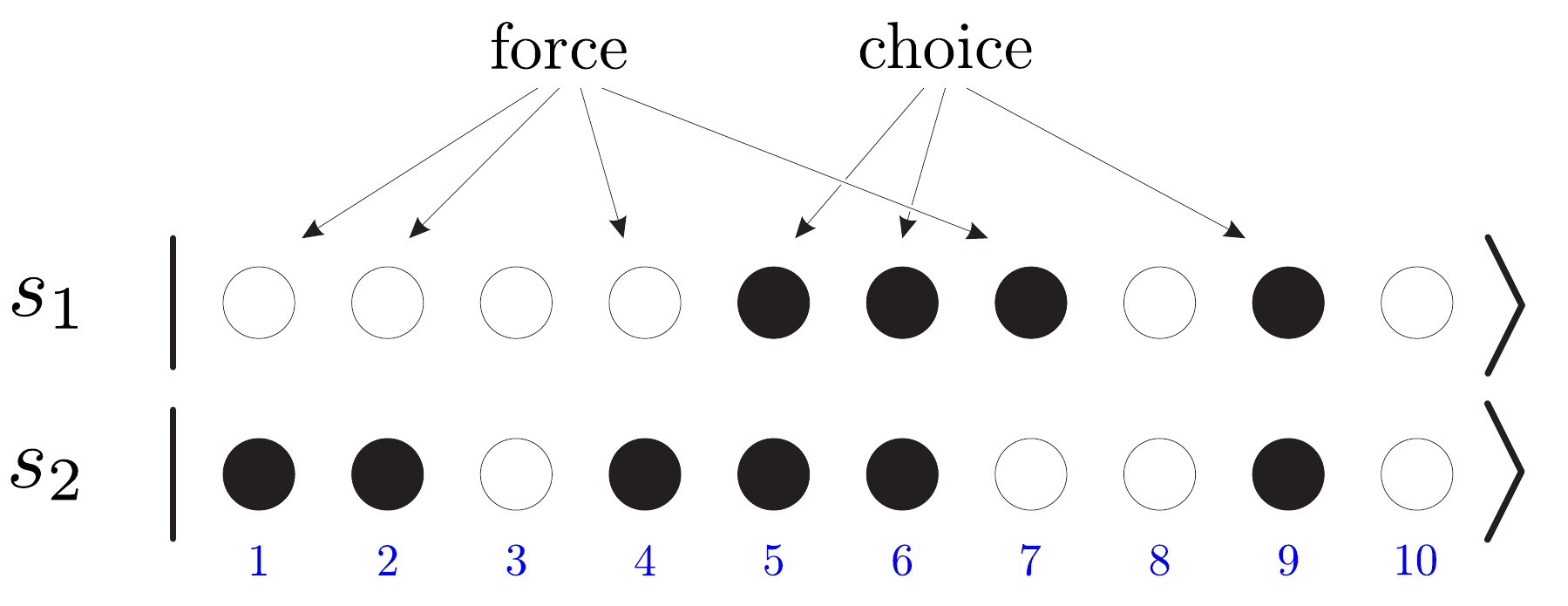}
\end{center}
\caption{Alignment of two Slater determinants for the annihilation operation. ``Force'' labels the orbitals which are either occupied in $s_1$ or $s_2$, but not in both, whereas ``choice'' labels all orbitals occupied in $s_1$ as well as $s_2$.}
\label{fig:RdmPair}
\end{figure}

The basic algorithmic idea is exemplified in figure~\ref{fig:RdmPair}. Namely, we subsume all orbitals occupied either in $s_1$ or $s_2$, but not in both, as ``force'' group, and all orbitals occupied in both $s_1$ and $s_2$ as ``choice'' group. The corresponding bit patterns $f_\mathrm{mask}$ and $c_\mathrm{mask}$ are computed in lines~\ref{algline:forcemask}~and~\ref{algline:choicemask} of Algorithm~\ref{alg:SlaterRDM} by a single bit operation. Since $\hprod{a_{t_2} s_2}{a_{t_1} s_1}$ is nonzero only if $a_{t_1} s_1 = \pm a_{t_2} s_2$, all occupied ``force'' orbitals have to be annihilated by $a_{t_1}$ and $a_{t_2}$, respectively. On the other hand, each ``choice'' orbital annihilated by $a_{t_1}$ must also be annihilated by $a_{t_2}$ and vice versa, but there's a freedom in exactly which of these orbitals to select, hence the ``choice'' designator. In our example, the only force orbital occupied in $s_1$ is $7$, so $t_1$ must contain $7$ but may ``choose'' between $5$, $6$ and $9$. If $p_1 = 3$, we obtain $t_1$ equal to one of $\ket{5\,6\,7}$, $\ket{5\,7\,9}$ or $\ket{6\,7\,9}$. The respective $t_2$ states are then $\ket{1\,2\,4\,5\,6}$, $\ket{1\,2\,4\,5\,9}$ and $\ket{1\,2\,4\,6\,9}$. After the obligatory annihilation sign factor calculations~\eqref{eq:AnnihilSignFactor}, the final result (for $p_1 = 3$) is
\begin{equation*}
\gamma_{\ket{5\,6\,7\,9}\bra{1\,2\,4\,5\,6\,9}} = \ket{5\,6\,7}\bra{1\,2\,4\,5\,6} - \ket{5\,7\,9}\bra{1\,2\,4\,5\,9} - \ket{6\,7\,9}\bra{1\,2\,4\,6\,9}.
\end{equation*}

Algorithm~\ref{alg:SlaterRDM} implements equation~\eqref{eq:AnnihilSignFactor} in line~\ref{algline:signfactor} and the first term of line~\ref{algline:appendresult}. $s_{\mathrm{force},k}$ stores the orbitals which must be annihilated in $s_k$ ($k = 1,2$), and the number of to-be annihilated ``choice'' orbitals in $s_1$ is computed in line~\ref{algline:numchoice1}. The while loop accumulates the return value list $r$ containing the ket-bra's as in the above example. In line~\ref{algline:bitdistribute}, the algorithm uses the '$\mathrm{BitDistribute}$' command, which basically just shifts bits to the positions designated by the 1-bits in $c_\mathrm{mask}$.
\begin{algorithm}
\caption{SlaterRDM}
\label{alg:SlaterRDM}
\begin{algorithmic}[1]
\REQUIRE $s_1$, $s_2$: \textbf{bitfield}, $p_1$: \textbf{int}
\STATE $f_\mathrm{mask} \Leftarrow s_1 \bitxor s_2$ \quad \label{algline:forcemask} \COMMENT{``force'' mask}
\STATE $s_{\mathrm{force},k} \Leftarrow \left(f_\mathrm{mask} \bitand s_k\right)$ ($k = 1,2$)
\STATE $n_{\mathrm{choice},1} \Leftarrow p_1 - \bitcount{s_{\mathrm{force},1}}$ \label{algline:numchoice1}
\IF{$n_{\mathrm{choice},1} < 0$}
	\RETURN 0
\ENDIF
\STATE $p_2 \Leftarrow \bitcount{s_2} - \bitcount{s_1} + p_1$
\STATE $a_{\mathrm{mask},k} \Leftarrow \mathrm{AnnihilSignMask}(s_k) \bitshiftleft 1$, $k = 1,2$
\STATE $\zeta \Leftarrow \prod_{k=1}^2 \revp{p_k} \cdot (-1)^\bitcount{a_{\mathrm{mask},k} \bitand s_{\mathrm{force},k}}$ \label{algline:signfactor} \COMMENT{sign factor}
\IF{$n_{\mathrm{choice},1} = 0$}
	\RETURN $\zeta \cdot \ket{s_{\mathrm{force},1}}\bra{s_{\mathrm{force},2}}$
\ENDIF
\STATE $c_\mathrm{mask} \Leftarrow s_1 \bitand s_2$ \quad \label{algline:choicemask} \COMMENT{``choice'' mask}
\STATE $k_\mathrm{choice} \Leftarrow \bitcount{c_\mathrm{mask}}$
\STATE $r \Leftarrow \{\}$
\STATE $t \Leftarrow \left(1 \bitshiftleft n_{\mathrm{choice},1}\right) - 1$
\WHILE[iterate Fermi map of 'choice' orbitals]{$(t \bitshiftright k_\mathrm{choice}) = 0$}
	\STATE $s_\mathrm{choice} \Leftarrow \mathrm{BitDistribute}\left(t, c_\mathrm{mask}\right)$ \label{algline:bitdistribute}
	\STATE append $r \Leftarrow \zeta \cdot \prod_{k=1}^2 (-1)^\bitcount{a_{\mathrm{mask},k} \bitand s_\mathrm{choice}} \ket{s_{\mathrm{force},1} + s_\mathrm{choice}}\bra{s_{\mathrm{force},2} + s_\mathrm{choice}}$ \label{algline:appendresult}
	\STATE $t \Leftarrow \mathrm{NextFermi}(t)$
\ENDWHILE
\RETURN r
\end{algorithmic}
\end{algorithm}

\subsection{Bosons}

\begin{figure}[ht!]
\begin{center}
\includegraphics[width=0.8\textwidth]{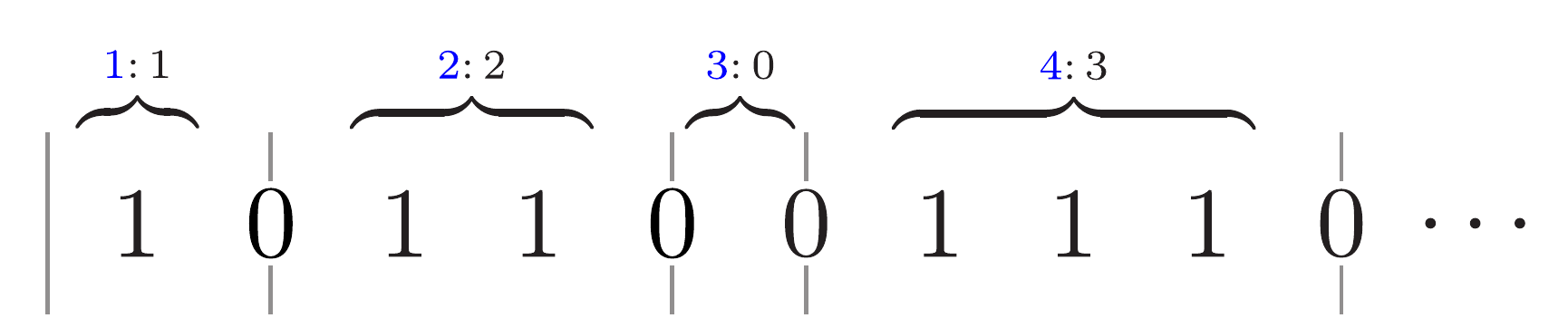}
\end{center}
\caption{Bit-encoding of a bosonic state. Blue numbers label modes, and 0-bits serve as delimiters between modes. The shown state consists of one boson in the 1st mode, two in the 2nd mode, zero in the 3rd and three in the 4th.}
\label{fig:BosonEncode}
\end{figure}

As a short outlook, we want to illustrate how the developed methods can easily be adapted to bosonic systems as well. In quantum mechanics, bosons are subatomic particles which obey Bose-Einstein statistics, like, for example, photons. For our purposes, we replace fermionic ``orbitals'' by bosonic ``modes'', which can be multiply occupied (i.e., the Pauli exclusion principle no longer holds for bosons). That is, the bosonic analogue of a fermionic Slater determinant differs only by the unrestricted number of particles in each mode. The central observation of this subsection states that a bit-encoding (equivalent to Slater determinants) works for bosons as well. The idea is detailed in figure~\ref{fig:BosonEncode}, where 0-bits serve as delimiters between modes. Lexicographical enumeration of bosonic states with a fixed total particle number $N$ and number of modes $m$ is accomplished via enumeration of the bit-encoded Slater determinants with $(m + N - 1)$ orbitals and $N$ particles! That is, Algorithm~\ref{alg:NextFermi} may be employed without modifications.

\section{Application to Transition Metal Atoms}
\label{sec:Application}

The application example is based on the series~\cite{NuclearChargeLimit2009,CI2009,Chromium2010}, in which~\cite{Chromium2010} makes use of the \texttt{FermiFab} toolbox to calculate ground state approximations for transition metal atoms (employing so-called configuration-interaction (CI) methods). The underlying quantum mechanical (non-relativistic, Born-Oppenheimer) Hamiltonian $H = H_0 + V_{ee}$ with
\begin{equation*}
H_0 = \sum_{i=1}^N \left(-\frac{1}{2} \Delta_{\vect{x}_i} - \frac{Z}{\abs{\vect{x}_i}}\right), \quad V_{ee} = \sum_{1 \le i < j \le N} \frac{1}{\abs{\vect{x}_i-\vect{x}_j}}
\label{eq:Hamiltonian}
\end{equation*}
governs atoms/ions with $N$ electrons and nuclear charge $Z$. The two terms in $H_0$ are the single-particle kinetic energy and nuclear potential, respectively, whereas the Coulomb operator $V_{ee}$ describes the pairwise inter-electron Coulomb repulsion. The Hamiltonian leaves the simultaneous eigenspaces of the well-known angular momentum, spin and parity ('LS') operators invariant, so calculating these eigenspaces \emph{first} leads to a huge dimension reduction. Specifically, the \texttt{FermiFab} toolbox automates the LS-eigenspace computation by combining configurations~\eqref{eq:ConfigTensor} with Clebsch-Gordan coefficients. We skip further details here; instead, for the purpose of this section, we provide two orthonormal LS-eigenstates of neutral Chromium ($N = Z = 24$) with symmetry level $^{5}D$:
{\small\begin{equation*}
\begin{split}
\psi_1 &:= \frac{1}{\sqrt{10}} \left(\vphantom{\sqrt{3}}\ket{3d_0\,3d_m\,3d_x\,4s\,\conj{4s}\,4d_x} - \ket{3d_0\,3d_m\,3d_y\,4s\,\conj{4s}\,4d_y} - \ket{3d_0\,3d_z\,3d_x\,4s\,\conj{4s}\,4d_y}\right.\\
&\left.-\ket{3d_0\,3d_z\,3d_y\,4s\,\conj{4s}\,4d_x} + \sqrt{3}\ket{3d_z\,3d_m\,3d_x\,4s\,\conj{4s}\,4d_y} - \sqrt{3}\ket{3d_z\,3d_m\,3d_y\,4s\,\conj{4s}\,4d_x}\right)
\end{split}
\end{equation*}}
and
{\small\begin{equation*}
\begin{split}
\psi_2 &:= \frac{1}{\sqrt{21}}\left(-\sqrt{3}/2\ket{3d_0\,4s\,\conj{4s}\,4p_z\,4p_x\,4d_y} - \sqrt{3}/2\ket{3d_0\,4s\,\conj{4s}\,4p_z\,4p_y\,4d_x}\right.\\
&+ 2 \ket{3d_m\,4s\,\conj{4s}\,4p_x\,4p_y\,4d_z} + \frac{1}{2}\ket{3d_m\,4s\,\conj{4s}\,4p_z\,4p_x\,4d_y} - \frac{1}{2}\ket{3d_m\,4s\,\conj{4s}\,4p_z\,4p_y\,4d_x}\\
&+ \frac{1}{2}\ket{3d_x\,4s\,\conj{4s}\,4p_x\,4p_y\,4d_y} + \ket{3d_x\,4s\,\conj{4s}\,4p_z\,4p_x\,4d_z} + \sqrt{3}\ket{3d_x\,4s\,\conj{4s}\,4p_z\,4p_y\,4d_0}\\
&- \ket{3d_x\,4s\,\conj{4s}\,4p_z\,4p_y\,4d_m} - \frac{1}{2}\ket{3d_y\,4s\,\conj{4s}\,4p_x\,4p_y\,4d_x} + \sqrt{3}\ket{3d_y\,4s\,\conj{4s}\,4p_z\,4p_x\,4d_0}\\
&+ \ket{3d_y\,4s\,\conj{4s}\,4p_z\,4p_x\,4d_m} + \ket{3d_y\,4s\,\conj{4s}\,4p_z\,4p_y\,4d_z} - 2\ket{3d_z\,4s\,\conj{4s}\,4p_x\,4p_y\,4d_m}\\
&+ \left.\,\frac{1}{2}\ket{3d_z\,4s\,\conj{4s}\,4p_z\,4p_x\,4d_x} + \frac{1}{2}\ket{3d_z\,4s\,\conj{4s}\,4p_z\,4p_y\,4d_y}\right).
\end{split}
\end{equation*}}
In this notation, $\conj{\,\cdot\,}$ means spin down $\downarrow$, otherwise up $\uparrow$, and the $s,p,d$ subshell orbitals are labeled $s$, $p_z\,p_x\,p_y$ and $d_0\,d_z\,d_m\,d_x\,d_y$, respectively. The numbers 3 and 4 denote the third and fourth shell. Since all spin-orbitals up to $3p$ are fully occupied, they are not shown here for conciseness of notation.

The following paragraph demonstrates how to translate the expectation value $\hprod{\psi_2}{V_{ee}\,\psi_1}$ into a list of Coulomb integral symbols
\begin{equation}
\label{eq:CoulombIntDef}
\bprod{a b}{c d} := \int_{\R^6} a^*(\vect{x}_1) b(\vect{x}_1)\,\frac{1}{\abs{\vect{x}_1-\vect{x}_2}}\,c^*(\vect{x}_2) d(\vect{x}_2) \ud\vect{x}_1 \ud\vect{x}_2,
\end{equation}
where $a,b,c,d \in L^2(\R^3)$ are \emph{spatial} orbitals and $^*$ denotes complex conjugation. As shown in~\eqref{eq:AvrRDM}, the essential step is the calculation of the 2-body reduced density matrix $\gamma_{\ket{\psi_1}\bra{\psi_2}}$. Using the Mathematica version of \texttt{FermiFab}, this is accomplished by the first line of the following code sample (see \texttt{mathematica/RDMdemo.nb}); the subsequent code just displays the result:\\
\includegraphics[width=\textwidth]{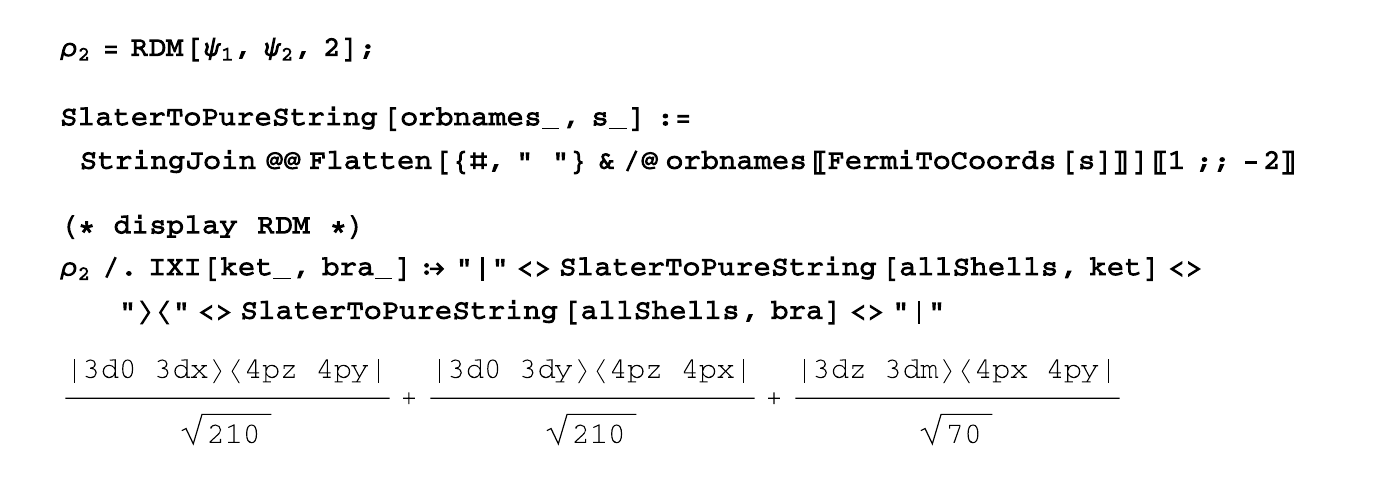}
The \texttt{FermiToCoords} command converts any bit-encoded Slater determinant to a vector of integers enumerating the occupied orbitals.

Since the Coulomb operator is independent of spin, we may effectively ``trace out'' the spin coordinate from the employed spin-orbitals. Specifically, consider single-particle wavefunctions
\begin{equation*}
\chi_i(\vect{x},s) = \varphi_i(\vect{x}) \alpha_i(s), \quad \vect{x} \in \R^3, s \in \left\{\uparrow, \downarrow\right\}, \quad i = 1,\dots,4
\end{equation*}
which factor into the spatial part $\varphi_i$ and spin part $\alpha_i$. Endowing particle $i$ with coordinates $(\vect{x}_i,s_i)$, the antisymmetrized $2$-body Slater determinants read
\begin{equation*}
\ket{\chi_i\,\chi_j} \equiv \frac{1}{\sqrt2} \left(\varphi_i(\vect{x}_1)\alpha_i(s_1)\, \varphi_j(\vect{x}_2)\alpha_j(s_2) - \varphi_j(\vect{x}_1)\alpha_j(s_1)\, \varphi_i(\vect{x}_2)\alpha_i(s_2)\right).
\end{equation*}
Plugged into the following equation for the Coulomb expectation value yields
\begin{equation*}
\label{eq:CoulombSpatialSpin}
\begin{split}
&\quad \hprod{\chi_1\,\chi_2}{\frac{1}{\abs{\vect{x}_1-\vect{x}_2}} \chi_3\,\chi_4}\\
&= \bprod{\varphi_1 \varphi_3}{\varphi_2 \varphi_4} \hprod{\alpha_1}{\alpha_3} \hprod{\alpha_2}{\alpha_4}\\
&- \bprod{\varphi_1 \varphi_4}{\varphi_2 \varphi_3} \hprod{\alpha_1}{\alpha_4} \hprod{\alpha_2}{\alpha_3}.
\end{split}
\end{equation*}
Translating this equation to alternating spin up $\uparrow$ and down $\downarrow$ orbitals (and taking symmetries of $\bprod{a b}{c d}$ into account) is accomplished by the \texttt{SpinTraceCoulomb} command in the first line of the following code sample:\\
\includegraphics[width=\textwidth]{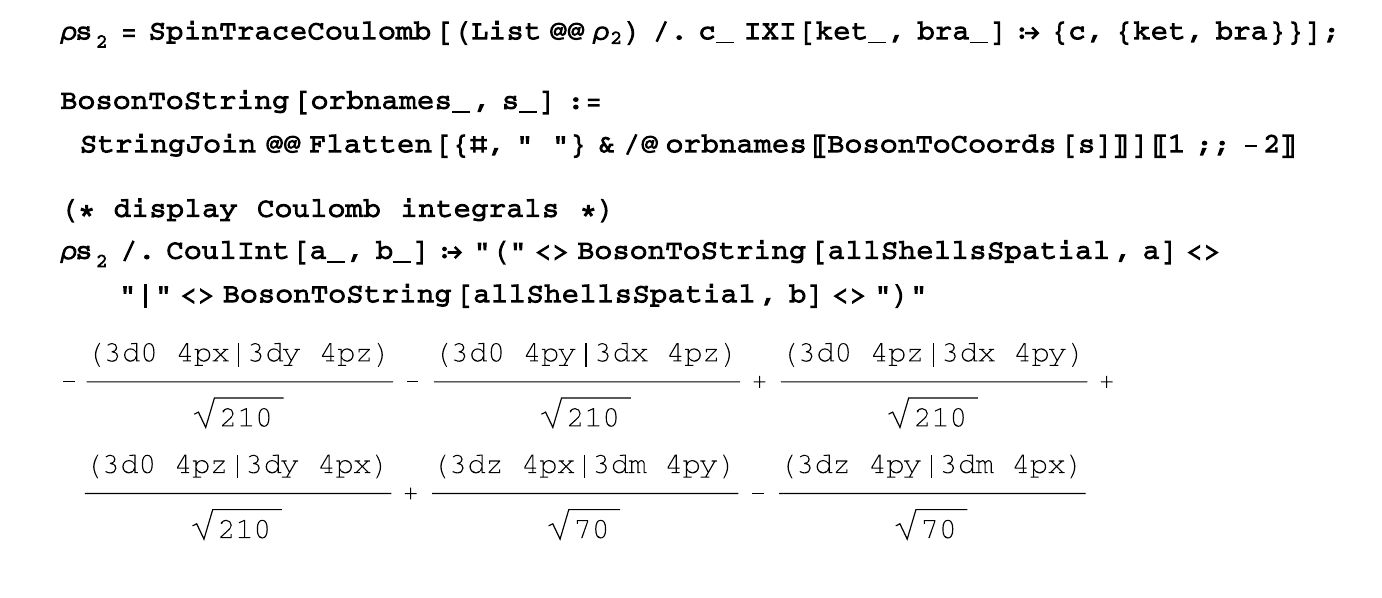}
Note that spatial orbitals can appear twice within a Coulomb integral symbol, e.g., $\bprod{a a}{b c}$. Thus, a bosonic encoding of these spatial orbitals is used to accommodate multiple occurrences, and hence the \texttt{BosonToCoords} command.

Concluding, we have obtained the desired list of Coulomb integral symbols, which may then be evaluated by inserting concrete functions into~\eqref{eq:CoulombIntDef}.

\section*{Acknowledgments}

I'd like to thank Gero Friesecke and Ben Goddard for many helpful discussions and feedback during the last two years. The toolbox inception was in spring 2008 for automating the (somewhat tedious) pen-and-paper calculations in~\cite{NuclearChargeLimit2009,CI2009}. These papers are concerned with the Schr\"odinger equation for atoms and ions from $N = 1$ to $10$ electrons. Specifically, \cite{NuclearChargeLimit2009,CI2009}~exploits the angular momentum, spin and parity symmetries of atoms to escape the prohibitively large dimensions incurred by quantum mechanical many-particle systems. Yet, application to atoms with even higher electron numbers ($\sim 30$) requires symbolic computer algebra. In~\cite{Chromium2010}, we specifically treat 3d transition metal atoms and use some algorithmic improvements incorporated into the \texttt{FermiFab} toolbox.

\bibliographystyle{elsarticle-num}

\end{document}